\documentclass[aps,pre,twocolumn,groupedaddress,showpacs]{revtex4}

\usepackage{graphicx} 
\usepackage{dcolumn} 
\usepackage{bm}
\newcommand{\ignore}[1]{} 
\usepackage{epsfig} 
\usepackage{color}
\usepackage{hyperref}
\usepackage{amsmath,amssymb}

\usepackage{comment}
\usepackage[TS1,T1]{fontenc}
\usepackage[latin1]{inputenc}

\begin{document}

\title{Disease and information spreading at different speeds in multiplex networks}

\author{F\'{a}tima Vel\'{a}squez-Rojas}
\affiliation{Instituto de F\'{i}sica de L\'{i}quidos y Sistemas Biol\'{o}gicos (UNLP-CONICET), 1900 La Plata, Argentina}
\author{Paulo Cesar Ventura}
\affiliation{Instituto de F\'{i}sica de  S\~{a}o Carlos, Universidade de S\~{a}o Paulo, S\~{a}o Carlos, SP, Brazil}
\author{Colm Connaughton}
\affiliation{Mathematics Institute, University of Warwick, Gibbet Hill Road, Coventry CV4 7AL, UK}
\affiliation{Centre for Complexity Science, University of Warwick, Coventry CV4 7AL, UK}
\author{Yamir Moreno}
\affiliation{Institute for Biocomputation and Physics of Complex Systems (BIFI), University of Zaragoza, 50018 Zaragoza, Spain}
\affiliation{Department of Theoretical Physics, University of Zaragoza, 50018 Zaragoza, Spain}
\affiliation{ISI Foundation, Turin, Italy}
\author{Francisco A. Rodrigues}
\affiliation{Instituto de Ci\^{e}ncias Matem\'{a}ticas e de Computa\c{c}\~{a}o, Universidade de S\~{a}o Paulo, S\~{a}o Carlos, S\~{a}o Paulo, Brazil}
\author{Federico Vazquez}
\affiliation{Instituto de C\'alculo, FCEN, Universidad de Buenos Aires and CONICET, Buenos Aires, Argentina}

\date{\today}

\begin{abstract}
Nowadays, one of the challenges we face when carrying out modeling of epidemic spreading is to develop methods to control disease transmission.  In this article we study how the spreading of knowledge of a disease affects the propagation of that disease in a population of interacting individuals.  For that, we analyze the interaction between two different processes on multiplex networks: the propagation of an epidemic using the susceptible-infected-susceptible dynamics and the dissemination of information about the disease --and its prevention methods-- using the unaware-aware-unaware dynamics, so that informed individuals are less likely to be infected.  Unlike previous related models where disease and information spread at the same time scale, we introduce here a parameter that controls the relative speed between the propagation of the two processes.  We study the behavior of this model using a mean-field approach that gives results in good agreement with Monte Carlo simulations on homogeneous complex networks.  We find that increasing the rate of information dissemination reduces the disease prevalence, as one may expect.  However, increasing the speed of the information process as compared to that of the epidemic process has the counter intuitive effect of increasing the disease prevalence.  This result opens an interesting discussion about the effects of information spreading on disease propagation.
\end{abstract}

\maketitle

\section{Introduction}
\label{sec:int}

Mathematical modeling of contagious disease spreading has become an important tool to estimate the extent of an epidemic \cite{Pastor-Satorras-2015,Arruda-2018}, and it is regaining attention with the actual coronavirus worldwide pandemic.  The knowledge or information we handle about a virus and its transmission among individuals plays a fundamental role in the containment of an epidemic.  This knowledge may lead to adopt strategies that change human behavior, with direct consequences on disease spreading, which has represented an intense research topic over the last years \cite{Pastor-Satorras-2015,Arruda-2018,Wang016,Lima-2015}.  It is known that the information about a disease and how this can contribute to epidemic spreading might help to develop more effective prevention methods \cite{funk2010modelling,manfredi2013modeling,Meloni011,Wang-2015-1,Funk_15}.
Some of these methods can significantly reduce the full extent of an epidemic, as shown in previous studies \cite{Funk6872,Granell-2013,Granell-2014}.  To explore the influence of human behavior on the spread of an epidemic, these works have used a model for the spreading of rumors to simulate the spread of knowledge about the disease (and its methods of prevention) by word of mouth.  In this way, the rumors --also called information-- and the epidemic are considered as two diffusion processes that interact with each other.  Some pioneer works on interacting spreading processes have already considered the dynamical interaction between two epidemics that propagate on single \cite{Goldenberg-2005,Newman-2005,Karrer-2011} and overlay \cite{Marceau-2011} networks.  Other more recent studies have also analyzed the impact of the information on the spread of epidemics in a population of interacting individuals \cite{Hang-Hyun-2006,Wang-2014,Wang-2016, guo2015cascade,wu_2012,FVR-2017,Col-Br-1}.

These systems with two interacting spreading dynamics can be studied using the topology of a multiplex network, where the disease and the information to prevent transmission spread in two different layers.  The disease layer may represent physical or proximity contacts for the spread of airborne diseases in people who interact regularly (family, coworkers, etc.) or occasionally (people who share public transport). The information layer represents contacts between people who exchange information face-to-face or in a virtual way by means of social networks.  To model the spreading of awareness (information) in this entangled epidemic-information processes, Granell et al. \cite{Granell-2013,Granell-2014} implemented the susceptible-infected-susceptible (SIS) dynamics, while Wang et al. \cite{Wang-2014,Wang-2016} used the susceptible-infected-recovered (SIR) dynamics.  In \cite{Granell-2013,Granell-2014} they showed that the degree of immunization of the informed individuals and the mass media change the critical aspects of disease spreading.  Besides, in \cite{Wang-2014,Wang-2016} the authors showed that there is an optimal information transmission rate that minimizes the disease spreading.  These works, however, assumed that the time scales associated to the propagation of the epidemic and the awareness processes are the same, while in principle one may expect that in real life  epidemics and information does not necessarily spread at the same speed.

In this context, we introduced in a recent article \cite{Col-Br-1} a new model of epidemic spreading with awareness considering the SIS dynamics for disease transmission and the dynamics of the Maki-Thompson rumor model \cite{Maki} for rumor dissemination. We also considered an external parameter $\pi$ that allows to control the relative timescales between the disease and rumor propagation processes.  A remarkable result of this model is that the prevalence of the disease increases with $\pi$, that is, as the transitions of the rumor process happen faster than those of the epidemic process.  This is a counter intuitive behavior, as one would  expect that a faster informational process should be more efficient in reducing the disease propagation and prevalence.  We note that in a previous work \cite{Karrer-2011} the authors studied a model for the interplay between two competing epidemics that propagate at different speeds, which are controlled by the time step $\Delta t$ of each process.  However, unlike the model studied by Ventura et. al. \cite{Col-Br-1} that uses an SIS-type dynamics on a two-layer network and individuals can be in the infected and informed states at the same time, the work in \cite{Karrer-2011} assumes that the two diseases spread on a single network following the susceptible-infected-recovered dynamics and that each individual can catch at most one of the two diseases (cross-immunity).

In this article we consider a simplified version of the model studied in \cite{Col-Br-1}, in order to understand the surprising influence of information awareness on the epidemic prevalence.  Our simulation results on multilayer networks turn to be qualitatively the same as those obtained in \cite{Col-Br-1}. However, we provide a continuous time formulation and a more complete theoretical analysis than performed before. We show that the resulting effects of varying the relative speed of infection and information processes are robust under models with a cyclic dynamics, which adds more evidence for the universal behavior of dynamical processes on multilayer networks \cite{Bianconi-2018}.  A mean-field (MF) approach helps to elucidate the mechanisms at play that give rise to some of the non-intuitive behavior mentioned above.  

The article is organized as follows: In section \ref{model}, we introduce the multiplex framework and the dynamics of the model on each layer.  We present numerical results in section \ref{results} and develop an analytical approach in section \ref{MF-approach}.  Finally, in section \ref{summary} we give a summary and conclusions.

\section{The Model}
\label{model}

\begin{figure}[t]
  \includegraphics[width=\columnwidth]{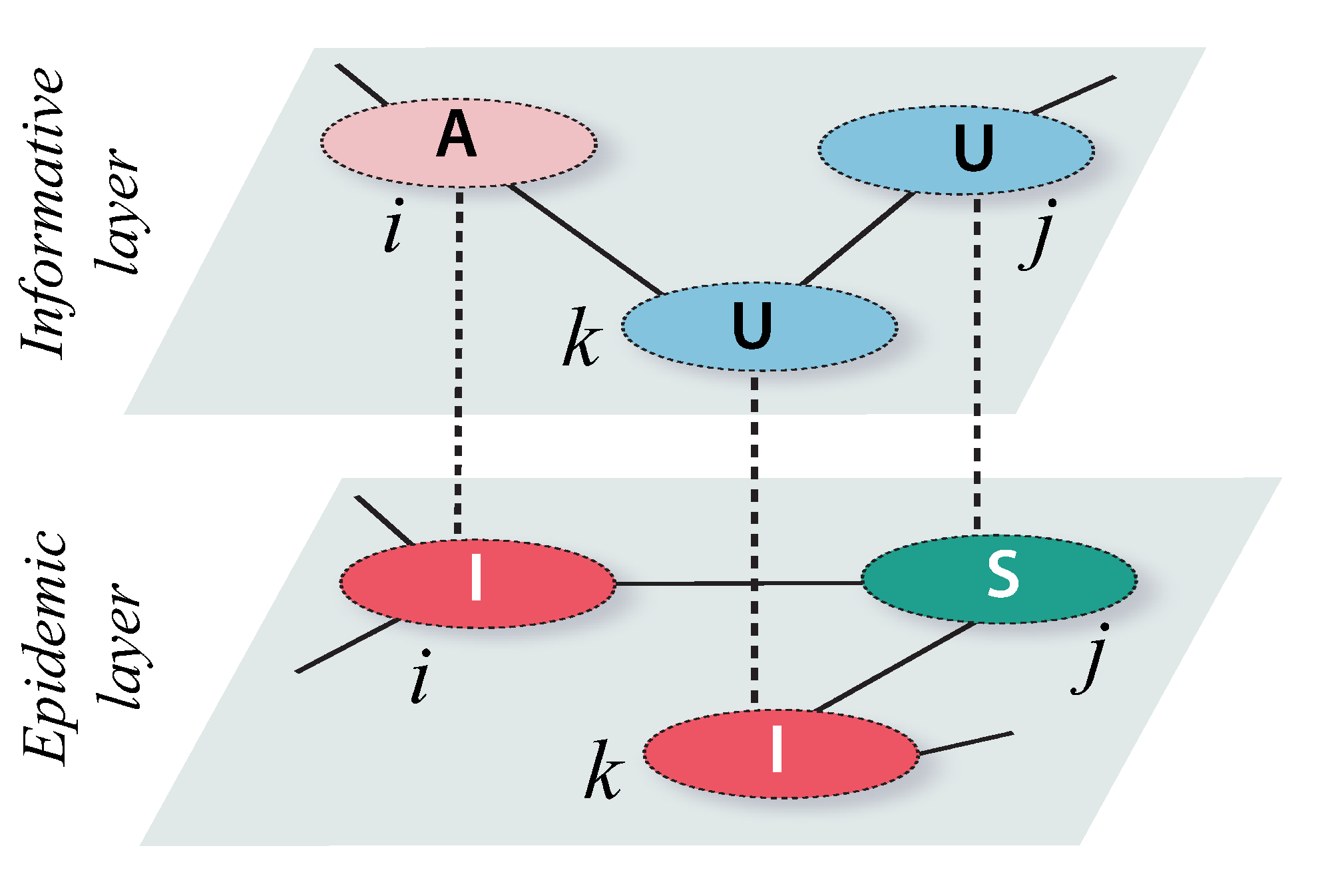}
  \caption{Schematic illustration of a multiplex structure used for the SIS-UAU model. In the information layer, nodes have two possible states: unaware (U) and aware (A) of the disease. In the epidemic layer, nodes represent the same individuals as in the top layer and can be either susceptible (S) or infected (I).}
  \label{two-layers}
\end{figure}

We consider a two-layer network made of an epidemic layer, where the disease propagates, and an information layer, where the disease awareness takes place, as shown in Fig.~\ref{two-layers}.  In the epidemic layer, nodes can be either \emph{Susceptible} (S) or \emph{Infected} (I), while in the information layer nodes are either in the \emph{Unaware} (U) state (an individual not aware of the disease) or in the \emph{Aware} (A) state (subjects who are aware of the disease). We represent the composite state of a node with two capital letters, the first one for the epidemic state and the second one for the information state, i.e., \emph{Susceptible--Unaware} (SU), \emph{Susceptible--Aware} (SA), \emph{Infected--Unaware} (IU), and \emph{Infected--Aware} (IA).

The basic SIS dynamics, in which infected nodes transmit the disease to susceptible neighbors with rate 
$\beta$ and recover from the disease at rate $\mu$, is modified to introduce the interaction between information and epidemics. The information is considered as the knowledge of the prevention methods that aware individuals have to reduce the probability of contracting the disease.  This is modeled as a reduction in the contagion rate by a factor $\Gamma$ ($0 \leq \Gamma \leq 1$) if the susceptible node is aware.  Then, an infected node infects an $SU$ neighbor with rate $\beta$, while the infection rate is reduced to 
$\Gamma \beta \le \beta$ if the neighbor is in the $SA$ state.  The dynamics on the information layer is quite similar to that of the SIS model, i.e., an unaware node becomes aware with rate $\gamma$ by contacting an aware neighbor, and aware nodes forget the information --or simply lose interest on it-- and go back to the unaware state at rate $\alpha$.  Besides, the existence of infected nodes reinforces the information about the disease, which is included in the model as a "self-awareness'' of the infected people, where IU nodes spontaneously become aware at rate $\kappa$.

\begin{figure}[t]
  \includegraphics[width=\columnwidth]{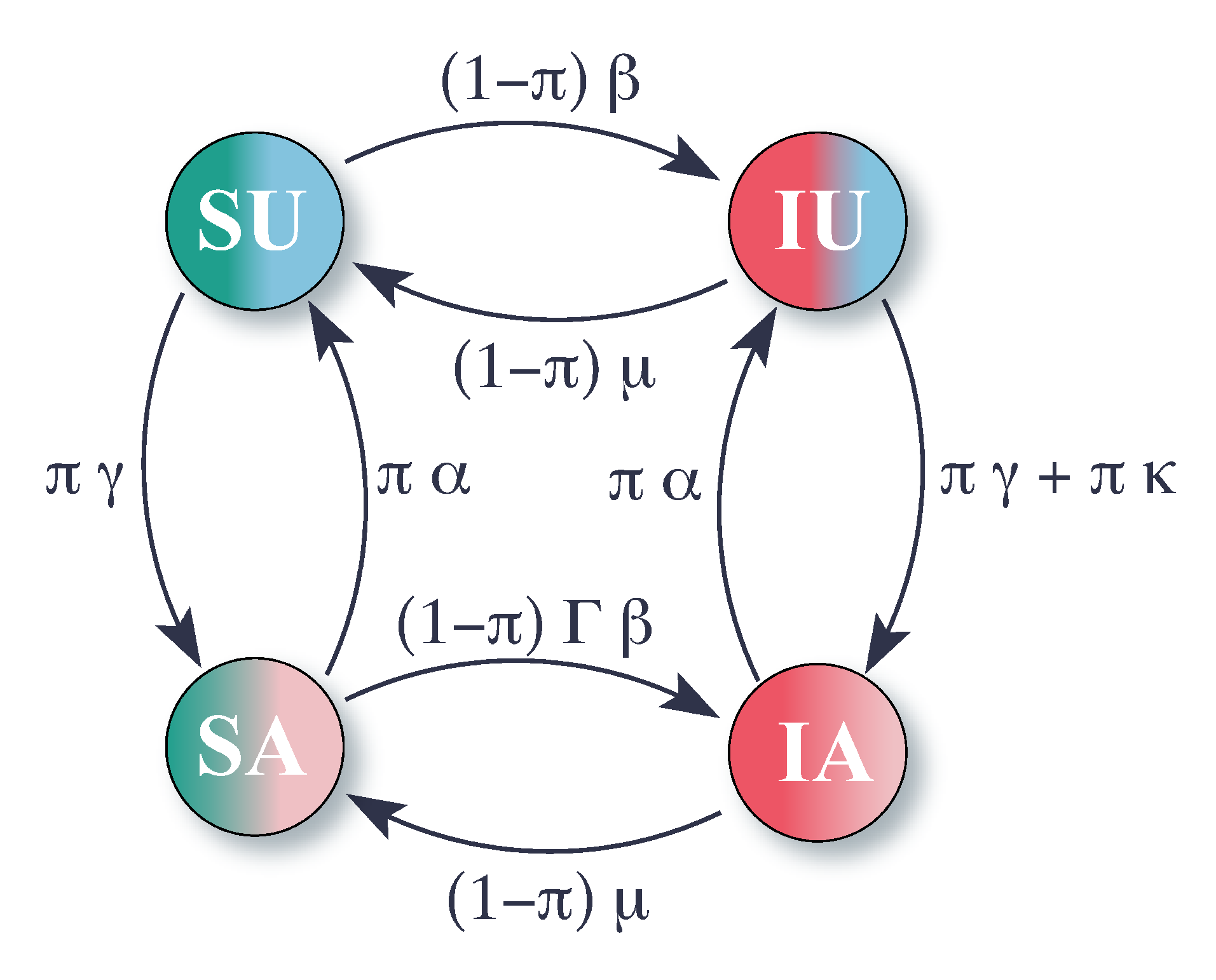}
  \caption{Schematic representation of the transitions between node states and their associated rates.}
  \label{state-transitions}
\end{figure}

As mentioned before, in real life it is expected that both the epidemic and information dynamics do not necessarily evolve at the same speed.  For this reason we introduce a parameter $\pi$ ($0 \le \pi \le 1$) that tunes the relative timescales associated with the disease and rumor propagation processes, by making the information and disease transitions proportional to $\pi$ and $(1-\pi)$, respectively.  That is, $\pi$ increases the speed of the information process as compared to the infection process, so that the final form of state transitions and their rates are:  
\begin{eqnarray*}
 Ix + SU & \xrightarrow{\text{\mbox{$(1-\pi)\beta$}}} & Ix + IU,  \\
 Ix + SA & \xrightarrow{\text{\mbox{$(1-\pi) \Gamma \beta$}}} & Ix + IA,  \\
 Ix & \xrightarrow{\text{\mbox{$(1-\pi) \mu$}}} & Sx,
\end{eqnarray*}
for the epidemic process, where $x=U,A$ represent an arbitrary information state, and 
\begin{eqnarray*}
 yU + yA & \xrightarrow{\text{\mbox{$\pi \gamma$}}} & yA + yA,  \\
 yA & \xrightarrow{\text{\mbox{$\pi \alpha$}}} & yU,  \\
 IU & \xrightarrow{\text{\mbox{$\pi \kappa$}}} & IA, 
\end{eqnarray*} 
for the information process, where $y=I,S$ represent an arbitrary epidemic state.   All these transitions are shown in Fig. \ref{state-transitions}.

\section{Numerical simulation results}
\label{results}

We perform numerical simulations of the model described in section~\ref{model} using a two-layer network made of two Erd\"{o}s-R\'enyi networks that represent the information and the epidemic layer, each one with $N=1000$ nodes and mean degree $\langle k \rangle=20$ (the typical number of different contacts per person reported in various surveys \cite{Hoang-2019}). The nodes in different layers represent the same individuals but their connections may differ in both layers. We analyze the behavior of the stationary density of infected nodes $\rho_i^*$ (disease prevalence) and the stationary density of aware nodes $\rho_a^*$.  We are particularly interested in studying how these two magnitudes are affected by the parameter $\pi$, which increases the speed of the information process as compared to that of the infection process.  

\begin{figure}[t]
  \includegraphics[width=\columnwidth]{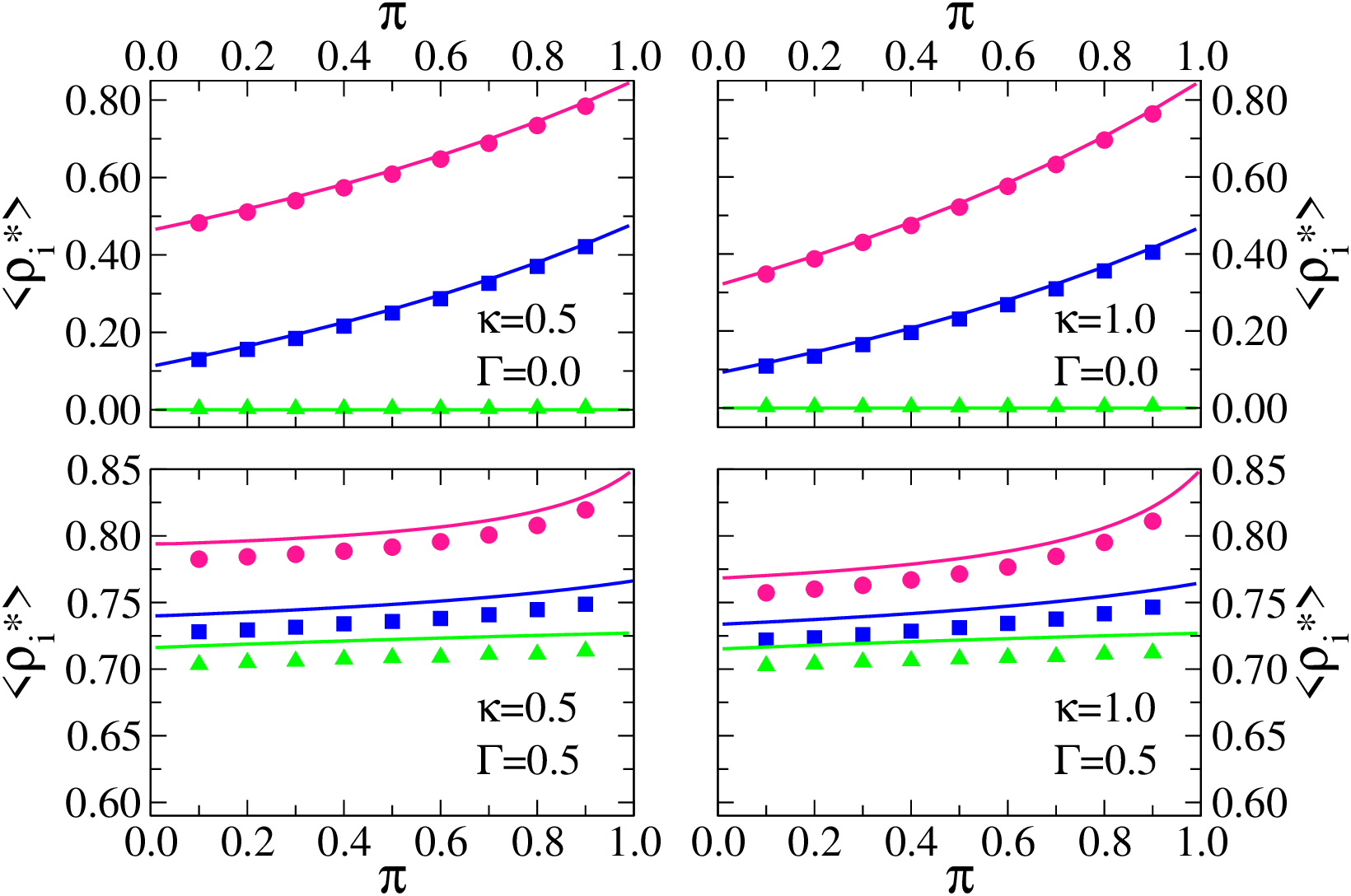}
  \caption{Average stationary density of infected nodes $\langle \rho_i^* \rangle$ vs information speed $\pi$, for $\gamma=0.0$ (circles), $0.1$ (squares) and $0.3$ (triangles), and for the values of $\kappa$ and $\Gamma$ indicated in each panel.  Other parameter values are $\beta=0.3$, $\mu=0.9$ and $\alpha = 0.6$.  Symbols correspond to MC simulation results while solid lines represent the analytical approximation, derived in section~\ref{MF-approach}.  The results are averaged over $10^4$ independent realizations of the spreading process starting from a density of infected nodes $\rho_i=0.5$ and aware nodes $\rho_a=0.5$ uniformly distributed over the epidemic and the information layer, respectively. Each layer is an Erd\"os-Renyi network of mean degree $\langle k \rangle=20$ and $N=1000$ nodes.}
  \label{rhoi-pi}
\end{figure}

In Fig.~\ref{rhoi-pi} we show simulation results for the average value of $\rho_i^*$ over $10^4$ independent realizations of the dynamics as a function of $\pi$, for various parameter values.  By comparing the top--left for panel with the bottom--left panel for $\kappa=0.5$, we notice that $\langle \rho_i^* \rangle$ is larger for $\Gamma=0.5$ than for $\Gamma=0$. We can see a similar behavior if we compare top--right and bottom--right panels for $\kappa=1$.  In general, we have verified that $\langle \rho_i^* \rangle$ increases as $\Gamma$ increases.  This is because the infection rate of $SA$ nodes increases with $\Gamma$, increasing the overall infection rate and so the disease prevalence.  The second and less intuitive result shown in this figure is that the prevalence increases monotonically with $\pi$ in all panels, which seems to be a quite robust behavior, independently on the parameter values.  Indeed, a similar behavior was also observed in our previous work \cite{Col-Br-1} using a more complex model, suggesting that this phenomenology may be universal in these type of models.  That is, speeding up the information dynamics with respect to the infection dynamics by increasing $\pi$, leads to a larger number of infected individuals at the stationary state.  This result result does not seem obvious given that we would expect that a faster information dynamics would be more efficient in reducing the number of infections.  By a "faster information dynamics" we mean that both information transmission and forgetting happen at higher rates, which are proportional to $\pi$.  In the next section we develop a MF approach that helps to elucidate this apparently contradictory result.

\begin{figure}[t]
  \includegraphics[width=\columnwidth]{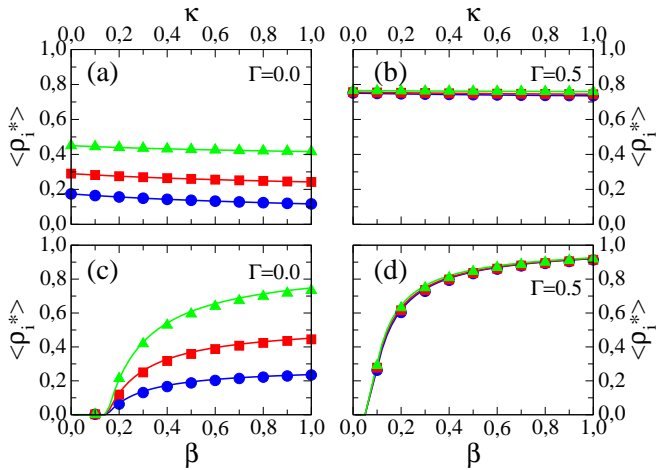}
  \caption{Top panels: $\langle \rho_i^* \rangle$ vs self-awareness rate $\kappa$ for $\gamma=0.1$, $\beta =0.3$, and (a) $\Gamma=0.0$ and (b) $\Gamma=0.5$.  Bottom panels: $\langle \rho_i^* \rangle$ vs infection rate $\beta$ for $\gamma=0.1$, $\kappa=0.5$, and (c) $\Gamma=0.0$ and (d) $\Gamma=0.5$.  Curves correspond to $\pi=0.1$ (circles), $0.5$ (squares) and $0.9$ (triangles).}
  \label{rhoi-beta-kappa}
\end{figure}

We also notice in Fig.~\ref{rhoi-pi} that the increase of the prevalence with $\pi$ is less pronounced for $\Gamma=0.5$, and we have verified that the curves become independent of $\pi$ for $\Gamma=1$.  When $\Gamma=1$, the infection and recovery rates $(1-\pi)\beta$ and $(1-\pi)\mu$, respectively, are the same for both, unaware and aware nodes.  Therefore, the dynamics becomes equivalent to that of the standard SIS model, with a stationary density of infected nodes in a MF set up given by the expression $\rho_i^*=\frac{\beta \eta -\mu}{\beta \eta}=0.85$, which is independent of $\pi$ because the infection and recovery rates are both proportional to $1-\pi$.  Here $\eta$ is the mean degree of the network (see section~\ref{MF-approach}).  For $\Gamma=0$ and $\gamma=0.3$ the prevalence vanishes for all $\pi$ values (triangles in top panels), and thus the system is reduced to a standard cyclic UAU dynamics akin to that of the SIS model, with transmission and recovery information rates $\gamma$ and $\alpha$, respectively, giving a stationary density of aware nodes in MF $\rho_{a}^*=\frac{\gamma \eta - \alpha}{\gamma \eta}=0.9$.

In Fig.~\ref{rhoi-beta-kappa} we show the behavior of the prevalence for two values of $\Gamma$ and three values of $\pi$, as indicated in the legends.  Panels (a) and (b) show the prevalence as a function of the self-awareness rate $\kappa$.  We observe that the prevalence decreases with $\kappa$, confirming that the self-awareness is an effective method in reducing disease propagation.  However, for $\Gamma=0.5$ the impact of $\kappa$ on the prevalence is very small, and also the prevalence is almost independent on $\pi$ [panel (b)].  Panels (c) and (d) show the prevalence as a function of the infection rate $\beta$.  As it happens in panel (b), the prevalence barely varies with $\pi$ for $\Gamma=0.5$ [panel (d)].  We also observe a transition from a healthy phase (epidemic extinction) to an endemic phase (epidemic propagation) at a threshold value $\beta_c$, which is reminiscent of that found in the SIS model.   

\begin{figure}[t]
  \includegraphics[width=\columnwidth]{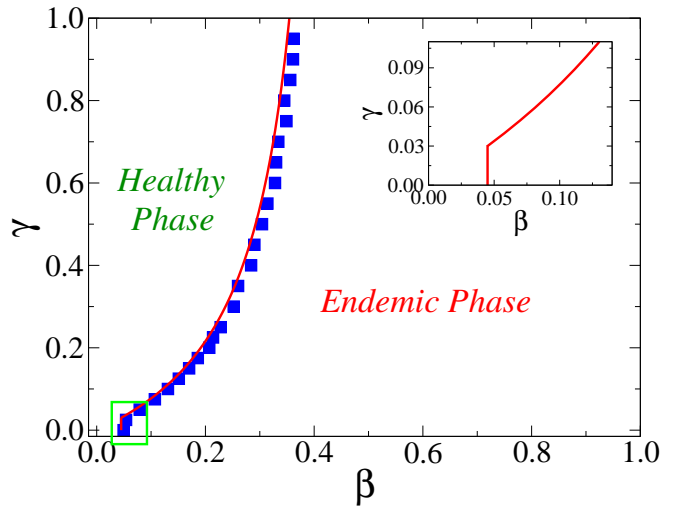}
  \caption{Phase diagram on the $\beta-\gamma$ plane showing the transition line between the healthy and endemic phases, for $\mu=0.9$, $\alpha = 0.6$, $\kappa=0.5$, $\Gamma=0.1$ and $\pi=0.5$.  Squares correspond to simulation results while the solid line represents the analytical approximation from Eq.~(\ref{bcrit}).  The inset is a zoom of the region indicated by a square, showing the analytical behavior of the transition line for small $\gamma$.}
  \label{df2d}
\end{figure}

To explore how the transition value $\beta_c$ depends on the information transmission rate $\gamma$, we calculated $\beta_c$ for $\pi=0.5$, $\Gamma=0.1$ and various values of $\gamma$ in the interval $(0,1)$.  Results are shown in the two-dimensional $\beta-\gamma$ phase diagram of Fig.~\ref{df2d}, where the square symbols represent the transition values that separate the healthy and endemic phases, calculated numerically.  For a given $\gamma$, we simulated the quasi-stationary state as proposed by Ferreira and others in \cite{Ferreira-2012-epidemic}, for several equally spaced values of $\beta$. The critical point $\beta_c$ was estimated as the value of $\beta$ that maximized the prevalence susceptibility, calculated as 
$\chi=N \left( \langle \rho_i^2 \rangle-\langle \rho_i \rangle^2 \right)/\langle \rho_i \rangle$, where $\langle \bullet \rangle$ represents an average over $1000$ independent realizations of the dynamics.  Starting from a population in the endemic phase with $\beta \lesssim 0.35$ and increasing $\gamma$ while keeping $\beta$ fixed, the system undergoes a transition to a healthy phase as $\gamma$ overcomes a threshold value $\gamma_c(\beta)$.  However, for $\beta \gtrsim 0.35$ the system remains in the endemic phase for all $\gamma$ values.  This means that, as long as the infection rate is low enough, the epidemics can be stopped by increasing the rate at which the information is transmitted between individuals but, strikingly, the information spreading is not able to stop the disease propagation when the infection rate is high enough.  

We also run simulations for other values of $\pi$ and $\Gamma$ (see Fig.~2 of the Supplementary Information).  These simulations reveal that the transition lines are independent of $\pi$.  Besides, the transition line $(\beta_c,\gamma_c)$ becomes more vertical as $\Gamma$ increases, until for $\Gamma=1.0$ it becomes the perfect vertical line $\beta_c \simeq 0.05$, independent of $\gamma$ and $\pi$.  An insight into these quite remarkable behaviors is given in section~\ref{MF-approach}.  

Summarizing the behavior of the model with respect to the parameters we can say that, on the one hand, the disease prevalence decreases when the information spreading rates increase through $\gamma$ and $\kappa$, or when the disease recovery rate $\mu$ increases.  On the other hand, the disease prevalence increases when the information recovery rate $\alpha$ decreases, or when the infection rate increases through $\beta$ and $\Gamma$.  These results are expected by model construction.  However, the prevalence increase with $\pi$ turns to be an unexpected and a striking result that seems harder to understand.  In section~\ref{MF-approach} we develop a MF approach that helps to gain an insight into these results.

\section{Mean-field approach}
\label{MF-approach}

We study the behavior of the SIS/UAU model using a mean-field approximation that assumes that, at every infinitesimal time step $dt$ of the dynamics, each node interacts with $\eta$ neighbors chosen at random among the nodes of the entire population (annealing approximation).  This approach neglects correlations that appear between the states of neighboring nodes in a static network, and should work reasonably well for random networks with homogeneous degree distributions and without degree correlations, such as the Erd\"{o}s-R\'enyi networks.  Then, the densities of nodes in each of the four states evolve according to the following set of coupled rate equations:
\begin{subequations}
  \begin{eqnarray}
    \label{drhodt-iu}
    \frac{d\rho_{iu}}{dt} &=& (1-\pi) \beta \eta \rho_{su} \rho_{i} + \pi
    \alpha \rho_{ia} - (1-\pi) \mu \rho_{iu} \nonumber \\ 
    &-& \pi \kappa \rho_{iu} - \pi \gamma \eta \rho_{iu} \rho_a, \\
    \label{drhodt-su}
    \frac{d\rho_{su}}{dt} &=& (1-\pi) \mu \rho_{iu} + \pi \alpha \rho_{sa}
    - (1-\pi) \beta \eta \rho_{su} \rho_i \nonumber \\ 
    &-& \pi \gamma \eta \rho_{su} \rho_a, \\
    \label{drhodt-ia}
    \frac{d\rho_{ia}}{dt} &=& \pi \gamma \eta \rho_{iu} \rho_a + \pi \kappa
    \rho_{iu} + (1-\pi) \Gamma \beta \eta \rho_{sa} \rho_i \nonumber \\ 
    &-& \pi \alpha \rho_{ia} - (1-\pi) \mu \rho_{ia},  \\
    \label{drhodt-sa}    
    \frac{d\rho_{sa}}{dt} &=& \pi \gamma \eta \rho_{su} \rho_a + (1-\pi)
    \mu \rho_{ia} - \pi \alpha \rho_{sa} \nonumber \\ 
    &-& (1-\pi) \Gamma \beta \eta \rho_{sa} \rho_i, 
  \end{eqnarray}
  \label{drhodt}  
\end{subequations}
where $\rho_{xy}$ is the density of nodes in state $xy$ ($x=i,s$ and $y=u,a$), 
$\rho_i=\rho_{iu}+\rho_{ia}$ is the density of infected nodes, and $\rho_a=\rho_{ia}+\rho_{sa}$ is the density of aware nodes. Also, the conservation relation for the total number of nodes $\rho_{iu}+\rho_{su}+\rho_{ia}+\rho_{sa}=\rho_i+\rho_s=\rho_a+\rho_u=1$ holds at any time.  The gain and loss terms of Eqs.~(\ref {drhodt}) correspond to the respective incoming and outgoing arrows at each of the four node states of Fig.~\ref {state-transitions}.  For instance, the gain term $(1-\pi) \beta \eta \rho_{su} \rho_i$ in Eq.~(\ref{drhodt-iu}) describes the fraction of nodes in state $SU$ that make the transition to state $IU$ per unit of time $dt$: an $SU$ node is infected at rate $(1-\pi) \beta$ by each of its infected neighbors, which are a total of $\eta \rho_i$ in average.

\subsection{Stationary states}
\label{stationary}

In this section we obtain solutions of the system of Eqs.~\ref{drhodt} at the stationary state.  We are particularly interested in the behavior of $\rho_i^*$ with $\pi$, which is the most intriguing as we showed in section~\ref{results}.  Given that Eqs.~\ref{drhodt} are a system of non-linear (quadratic) equations, explicit formulas for its stationary solutions can not be obtained with standard methods.  Therefore, it is hard to obtain closed expressions for the densities as a function of the parameters.  Instead, we derive here parametric equations that relate $\rho_i^*$ and $\pi$ through $\rho_a^*$ (the "parameter"), which is an indirect form of expressing $\rho_i^*$ as a function of $\pi$.   For that, we obtain expressions for the different stationary densities $\rho_{iu}^*$, $\rho_i^*$,  
$\rho_{su}^*$ and $\rho_{sa}^*$ as a function of $\rho_a^*$, as we show bellow.  

We start by adding Eqs.~(\ref{drhodt}a) and (\ref{drhodt}c) on one side, and Eqs.~(\ref{drhodt}c) and (\ref{drhodt}d) on the other side, to arrive to the following rate equations for $\rho_i$ and 
$\rho_a$, respectively:
\begin{subequations}
\begin{eqnarray}
\frac{d \rho_i}{dt} &=& (1-\pi) \left[ \beta \eta \left(\rho_{su}+\Gamma \rho_{sa} \right) - \mu \right] \rho_i, \\
\label{drhoadt}
\frac{d\rho_a}{dt} &=& \pi \left[ \gamma \eta (1-\rho_a)-\alpha \right] \rho_a + \pi \kappa \rho_{iu}.
\end{eqnarray}
\label{drira-dt}
\end{subequations}
A simple stationary solution of Eqs.~(\ref{drira-dt}) is obtained by setting   $\rho_i=0$, which leads to $\left[ \gamma \eta (1-\rho_a^*)-\alpha \right] \rho_a^*=0$ for $\pi \neq 0$.  Therefore, there are two trivial stationary states corresponding to a totally healthy population ($\rho_{iu}^*=\rho_{ia}^*=0,\rho_s^*=1$) in which (a) either all individuals are unaware ($\rho_{sa}^*=0$, $\rho_{su}^*=1$), or (b) there is a fraction $\rho_{sa}^*=\frac{\gamma \eta - \alpha}{\gamma \eta}$ of aware individuals.  This scenario corresponds to a simple UAU dynamics.  At the non-trivial stationary state $\rho_i^* \ne 0$, with $\pi \in (0,1)$, we obtain the equations
\begin{subequations}
\begin{eqnarray}
\label{dri}
\beta \eta \left(\rho_{su}^*+\Gamma \rho_{sa}^* \right) - \mu &=& 0, ~~ \mbox{and} \\
\label{dra}
\left[ \gamma \eta (1-\rho_a^*)-\alpha \right] \rho_a^* + \kappa \rho_{iu}^* &=& 0. 
\end{eqnarray}
\end{subequations}
Using the identities $\rho_{su}^*+\rho_{iu}^*=\rho_u^*=1-\rho_a^*$, $\rho_{sa}^*+\rho_{ia}^*=\rho_a^*$ and $\rho_i^*=\rho_{iu}^*+\rho_{ia}^*$ we can express $\rho_{su}^*$ and $\rho_{sa}^*$ in terms of $\rho_i^*$, $\rho_a^*$ and $\rho_{iu}^*$ as
\begin{subequations}
  \begin{eqnarray}
    \label{rsu}
    \rho_{su}^*&=&1-\rho_a^*-\rho_{iu}^* ~~~ \mbox{and} \\ 
    \label{rsa}
    \rho_{sa}^*&=&\rho_a^*- \rho_i^* + \rho_{iu}^*.   
  \end{eqnarray}
  \label{rsu-rsa}
\end{subequations}
Substituting the expressions Eqs.~(\ref{rsu-rsa}) for $\rho_{su}^*$ and $\rho_{sa}^*$ into Eq.~(\ref{dri}) and solving for $\rho_i^*$ we arrive to
\begin{equation}
  \rho_i^* = \frac{\beta \eta - \mu}{\Gamma \beta \eta} - \frac{(1-\Gamma) (\rho_a^*+\rho_{iu}^*)}{\Gamma}.
  \label{ri}
\end{equation}
Finally, replacing the expression 
\begin{eqnarray}
  \rho_{iu}^* = \frac{\left[ \alpha - \gamma \eta (1-\rho_a^*) \right] \rho_a^*}{\kappa}    
  \label{riu}
\end{eqnarray}
for $\rho_{iu}^*$ from Eq.~(\ref{dra}) into Eq.~(\ref{ri}) we obtain, after doing some algebra, the following equation that relates $\rho_i^*$ with $\rho_a^*$    
\begin{equation}
  \rho_i^* = \frac{\beta \eta - \mu}{\Gamma \beta \eta} - \frac{(1-\Gamma) \left[ \kappa + \alpha - \gamma \eta (1-\rho_a^*) \right] \rho_a^*}{\Gamma \kappa}.    
  \label{ri-1}
\end{equation}
We can also express $\rho_{su}^*$ and $\rho_{sa}^*$ in terms of $\rho_a^*$.  Inserting expression Eq.~(\ref{riu}) for $\rho_{iu}^*$ into Eq.~(\ref{rsu}) we arrive to
\begin{equation}
  \rho_{su}^* = 1- \frac{\left[ \kappa + \alpha - \gamma \eta (1-\rho_a^*) \right] \rho_a^*}{\kappa}.    
  \label{rsu-1}
\end{equation}
Then, replacing Eqs.~(\ref{riu}) and (\ref{ri-1}) for $\rho_{iu}^*$ and $\rho_{i}^*$, respectively, into Eq.~(\ref{rsa}) we obtain
\begin{equation}
  \rho_{sa}^* = \frac{\left[ \kappa + \alpha - \gamma \eta (1-\rho_a^*) \right] \rho_a^*}{\Gamma \kappa} - \frac{\beta \eta - \mu}{\Gamma \beta \eta}.    
  \label{rsa-1}
\end{equation}
Now that we have explicit expressions for the stationary densities $\rho_{iu}^*$, $\rho_i^*$, $\rho_{su}^*$ and $\rho_{sa}^*$ in terms of $\rho_a^*$ given by Eqs.~(\ref{riu}), (\ref{ri-1}), (\ref{rsu-1}) and (\ref{rsa-1}), respectively, we can obtain an expression that relates $\pi$ with $\rho_a^*$ by inserting these expressions into Eq.~(\ref{drhodt-su}) at the stationary state
\begin{equation}
    (1-\pi) \mu \rho_{iu}^* + \pi \alpha \rho_{sa}^* - \left[ (1-\pi) \beta \rho_i^* + \pi \gamma \rho_a^* \right] \eta \rho_{su}^* = 0,
\label{rhoa-pi-stat}
\end{equation}
and solving for $\pi$.  After doing some algebra, we finally obtain the following equation that gives $\pi$ as a function of the density $\rho_a^*$ and the other parameters:
\begin{equation}
  \pi = \frac{P(\rho_a^*)}{Q(\rho_a^*)},
  \label{pi-poly}
\end{equation}
where $P$ and $Q$ are polynomial of degree two and four in $\rho_a^*$ given by Eqs.~(\ref{P}) and (\ref{Q}), respectively, of Appendix \ref{completepoly}.  In principle, it is possible to transform Eq.~(\ref{pi-poly}) into a quartic equation in $\rho_a^*$ and find its solution, which would give an expression for $\rho_a^*$ as a function of the model's parameters and also an expression for $\rho_i^*$ by inserting this expression for $\rho_a^*$ into Eq.~(\ref{ri-1}).  However, as we can guess, the resulting expression would be highly complicated and not very useful.  Instead, we prefer to state the analytical relationship between $\rho_i^*$ and $\pi$ in the parametric form $[\pi(\rho_a^*),\rho_i^*(\rho_a^*)]$, where the expressions for $\pi(\rho_a^*)$ and $\rho_i^*(\rho_a^*)$ are given by Eqs.~(\ref{pi-poly}) and (\ref{ri-1}), respectively.  This parametric solution is plotted by solid lines in Fig.~\ref{rhoi-pi} and compared with MC simulation results (symbols).  We observe that the agreement between theory and simulations is quite good for $\Gamma=0$, but some discrepancies arise for $\Gamma=0.5$.

Even though the analytical solution presented above describes numerical data rather well, its complicated form makes it hard to explore the behavior of the densities with $\pi$.  Instead, to gain an insight into the behavior of $\rho_i^*$ with $\pi$ it proves useful to analyze the simplest non-trivial case $\gamma=0$ and $\Gamma=0$, where $\rho_i^*$ also exhibits the monotonic increase with $\pi$ observed for the general case $\gamma \ne 0$ and $\Gamma \ne 0$.  As we show in Appendix~\ref{gG0}, the stationary density of infected nodes for $\gamma=\Gamma=0$ adopts the rather simple form 
\begin{equation}
\rho_{i}^* = \frac{\alpha (\beta \eta - \mu) \left[\pi (\kappa+\alpha) +
    (1-\pi)\mu \right]}{(\kappa+\alpha) \beta \eta \left[ \pi \alpha +
    (1-\pi) \mu \right]}.
\label{rhoi-stat}
\end{equation}
We can check from expression Eq.~(\ref{rhoi-stat}) that for $\kappa=0$ is $\rho_i^*=\frac{\beta \eta -\mu}{\beta \eta}$, which corresponds to the stationary value of $\rho_i$ in the SIS model.  Indeed, when $\kappa=0$ and $\gamma=0$ there are no transitions to aware states $SA$ and $IA$, and thus all nodes are unaware at the steady state ($\rho_{su}^*+\rho_{iu}^*=1$), and subject to the standard SIS dynamics.  For $\kappa>0$, the term $\pi(\kappa+\alpha)$ in the numerator of Eq.~(\ref{rhoi-stat}) grows faster than the term $\pi \alpha$ in the denominator as $\pi$ increases, and thus $\rho_i^*$ increases when $\pi$ increases, as we have seen already for all parameter values analyzed in section~\ref{results}.

This result can be understood intuitively with the help of Fig.~\ref{state-transitions}, by analyzing the stationary flow between states.  On the one hand, we expect that $\rho_{sa}^*$ decreases as 
$\pi$ increases.  This is because the incoming flow $F_{ia \to sa} = (1-\pi)\mu \rho_{ia}^*$ (from $IA$ to $SA$ ) decreases with $\pi$, while the outgoing flow $F_{sa \to su} = \pi \alpha \rho_{sa}^*$ (from $SA$ to $SU$) increases with $\pi$.  On the other hand, we proved in Appendix~\ref{gG0} that 
$\rho_{su}^*$ is independent of $\pi$ and given by the expression
\begin{eqnarray}
\rho_{su}^* = \frac{\mu}{\beta \eta}.
\label{rhosu-stat}
\end{eqnarray}
Therefore, when $\pi$ increases the density of susceptible nodes $\rho_s^*=\rho_{su}^*+\rho_{sa}^*$ decreases, and thus $\rho_i^*$ increases.

It proves instructive to derive Eq.~(\ref{rhosu-stat}) from the analysis of the flows of Fig.~\ref{state-transitions}.  Given that in the steady state the incoming and outgoing flows in any node state is the same, we have that $F_{ia \to sa} = F_{sa \to su}$, and thus we can think that there is a net flow from $IA$ to $SU$ equal to
\begin{eqnarray}
F_{ia \to su}=(1-\pi)\mu \rho_{ia}^*.
\end{eqnarray}
Therefore, the total incoming flow to $SU$ from infected states is
\begin{eqnarray}
F_{i \to su} &=& F_{iu \to su} + F_{ia \to su}  \\
&=&(1-\pi) \mu \rho_{iu}^* + (1-\pi) \mu \rho_{ia}^* = (1-\pi)  \mu \rho_i^*, \nonumber 
\end{eqnarray}
while the outgoing flow from $SU$ to infected nodes is 
\begin{eqnarray}
F_{su \to i} = F_{su \to iu}=(1-\pi) \beta \eta \rho_{su}^* \rho_{i}^*.   
\end{eqnarray}
Then, the dynamics of the system corresponds to that of an SU $\to$ I $\to$ SU model, where we know that the stationary density of $SU$ nodes equals the ratio between the recovery rate $(1-\pi) \mu$ and the infection rate$(1-\pi)\beta \eta$, leading to Eq.~(\ref{rhosu-stat}).

\subsection{Stability analysis}
\label{stability-analysis}

A relevant feature in models of epidemic and information spreading is the existence of a transition from a healthy phase ($\rho_i^*=0$) to an endemic phase ($\rho_i^* >0$) as the infection probability overcomes a threshold value $\beta_c$, as we described in section~\ref{results} and showed in Figs.~\ref{rhoi-beta-kappa} and \ref{df2d}.  We want to find an analytical expression for the transition line $\beta_c(\gamma)$ of Fig. \ref{df2d}, along which the stability of the the healthy phase  changes, so that it is stable for $\beta < \beta_c$ and unstable for $\beta > \beta_c$.  For that, we perform a linear stability analysis of the stable fixed points within the healthy phase, which are 
\begin{eqnarray}
\vec{\rho}_1^{\,*}&=&(0,0,0,1) ~~~~~~~~~~~~~~~~ \mbox{for $\gamma \eta < \alpha$ and} \nonumber \\  
\vec{\rho}_2^{\,*}&=&\left( 0,\frac{\gamma \eta - \alpha}{\gamma \eta},0,\frac{\alpha}{\gamma \eta} \right) 
~~~ \mbox{for $\gamma \eta > \alpha$}.
\label{rho1rho2}
\end{eqnarray}
where $\vec{\rho}_n^{\,*} \equiv (\rho_{iu}^*,\rho_{sa}^*,\rho_{ia}^*,\rho_{su}^*)$, with $n=1,2$.  These are the two fixed points corresponding to the healthy phase obtained in section~\ref{stationary}, where the dynamics of aware nodes is given by Eq.~(\ref{drhoadt}) with $\rho_{iu}=0$
\begin{eqnarray*}
\frac{d\rho_a}{dt} =\pi \left[ \gamma \eta (1-\rho_a)-\alpha \right] \rho_a.
\end{eqnarray*}
The linearized form of this equation around $\rho_a=0$ corresponding to the fixed point $\vec{\rho}_1^{\,*}$ is $d\rho_a/dt = \lambda \rho_a$, with $\lambda \equiv \pi (\gamma \eta - \alpha)$.  Then, $\vec{\rho}_1^{\,*}$ is stable (unstable) for $\lambda <0$ ($\lambda >0$), as stated in Eqs.~(\ref{rho1rho2}) assuming $\pi \neq 0$.

\begin{figure}[t]
  \includegraphics[width=\columnwidth]{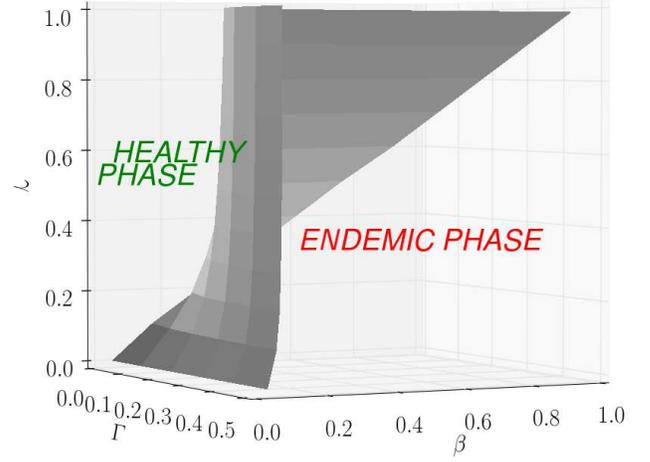}
  \caption{Phase diagram on the $\beta-\gamma-\Gamma$ space obtained from Eq.~(\ref{bcrit}) for the same parameter values as in Fig.~\ref{df2d}.}
  \label{df3d}
\end{figure}

In Appendix~\ref{stability} we perform a linear stability analysis of the fixed points $\vec{\rho}_n^{\, *}=(0,A,0,1-A)$, where 
\begin{eqnarray}
A &=& 0 ~~~~~~~~~~~ \mbox{for $\gamma \eta < \alpha$ ($n=1$) and} \nonumber \\ 
A &=& \frac{\gamma \eta -\alpha}{\gamma \eta} ~~~ \mbox{for $\gamma \eta > \alpha$ ($n=2$)}, 
\label{A}
\end{eqnarray}
and show that the following relation must hold at the transition point:
\begin{eqnarray}
\left[ (1-\pi)\mu+\pi(\gamma\eta+\kappa) \right] \left[ (1-\Gamma)\beta\eta A + \mu- \beta\eta \right] = 0.
\label{relation}
\end{eqnarray}
Given that we considered the rates $\mu, \gamma$ and $\kappa$ to be positive in simulations, the first term in brackets of Eq.~(\ref{relation}) is positive, thus we have 
\begin{eqnarray*}
(1-\Gamma) \beta \eta A +\mu-\beta\eta &=& 0. 
\end{eqnarray*}
Replacing the values of $A$ from Eqs.~(\ref{A}), we finally obtain the following expression for the critical infection rate:
\begin{eqnarray}
\beta_c =
\begin{cases}
\frac{\mu}{\eta}  ~~~ & \mbox{for $\gamma\eta < \alpha$ and} \\
\frac{\gamma \mu}{\gamma \eta - \left(1-\Gamma \right) \left(\gamma \eta- \alpha \right)} ~~~ & 
\mbox{for $\gamma\eta > \alpha$.}  
\end{cases}
\label{bcrit}
\end{eqnarray}
In Fig.~\ref{df2d} we observe that the analytical approximation of the transition line 
$\beta_c(\gamma)$ from Eq.~(\ref{bcrit}) (solid line) agrees quite well with the transition points obtained from simulations (squares).  We can also check that $\beta_c$ approaches the value $\mu/(\Gamma \eta) = 0.45$ in the $\gamma \to \infty$ limit (for $\mu=0.9, \Gamma=0.1$ and $\eta=20$), confirming that for high enough values of $\beta$ the information is not able to stop the epidemics, as mentioned in section~\ref{results}.  We also see that for $\Gamma=1$ is $\beta_c=\mu/\eta=0.045$ for all $\gamma$, which is in agreement with MC results (see section~I of the Supplementary Information).  Given that performing numerical simulations for various values of $\gamma$ and $\Gamma$ are very costly, we also implemented Eq.~(\ref{bcrit}) to build a transition plane in the $\beta-\gamma-\Gamma$ space.  Results are shown in the phase diagram of Fig.~\ref{df3d}.

\section{Conclusions} 
\label{summary}

We have explored the interplay between the propagation of an epidemic disease using the susceptible-infected-susceptible dynamics and the dissemination of information about the knowledge of the disease using the unaware-aware-unaware dynamics, as a simplified model from a recent study \cite{Col-Br-1}.  For that, we assumed that the disease and the information spread on two coupled Erd\"{o}s-R\'enyi networks where these two processes interact with each other, and whose relative propagation speeds are controlled by an external parameter $\pi$.  We have verified that the information helps to reduce the disease prevalence and increase the epidemic threshold of the disease. We have also observed that self-awareness, which keeps infected individuals aware of their condition, is a very effective mechanism for reducing the disease prevalence.  Surprisingly, the prevalence increases with $\pi$, that is, as the information dynamics is faster.  This seemingly counter intuitive result was also obtained in a more complex model studied in our previous work \cite{Col-Br-1} and, therefore, it seems to be universal and independent of the model details.  However, it was not fully explored and understood.   

In order to gain an insight into this phenomenon, we developed a MF approach to study the dynamics of the model.   We found a good agreement between simulations of the model and analytical MF results.  We showed that the SIS/UAU dynamics in MF exhibits a behavior that is qualitatively the same to that found in the SIS/UAU and SIS/UARU models using the Markov chain approach and Monte Carlo simulations \cite{Col-Br-1}, in particular, the increase of the prevalence with $\pi$.  Besides, the MF approach allowed for the detailed study of a simple non-trivial case where the relation between the prevalence and $\pi$ was analyzed in terms of probability flows between states.

It is interesting to note that the non-trivial relation between disease propagation and information spreading described in this article calls for a careful analysis of the impact of information management on disease spreading in a real society, something very pertinent in the current global pandemic.  Given that these results seem to hold for cyclic (SIS-like) spreading dynamics, both for disease and information processes, it would be worthwhile to explore whether a similar phenomena is observed in models where two non-cyclic (SIR-type) dynamics interact, with controllable relative speeds.  It might also be worth studying the behavior of the model on multilayer networks with more complex topologies than the Erd\"os-Renyi networks used in this work, such as scale-free networks or contact networks with a structure obtained from real data.

\acknowledgments
Y. M. acknowledges support from Intesa Sanpaolo Innovation Center, from the Government of Arag\'on and FEDER funds, Spain through grant ER$36-20$R to FENOL, and by MINECO and FEDER funds (grant FIS$2017-87519-$P). The funders had no role in study design, data collection, and analysis, decision to publish, or preparation of the manuscript. F.V. acknowledges financial support from CONICET (Grant No. PIP $0443/2014$) and from Agencia Nacional de Promoci\'on Cien\'itfica y Tecnol\'ogica (Grant No. PICT $2016$ Nro $201 0215$).  PCV thanks FAPESP for the PhD grant $2016/24555-0$. Research carried out using the computational resources of the Center for Mathematical Sciences Applied to Industry (CeMEAI) funded by FAPESP (grant $2013/07375-0$). FAR acknowledges financial support from the Conselho Nacional de Desenvolvimento Cient\'fico e Tecnol\'ogico (CNPq, Grant number $309266/2019-0$).

\appendix 

\section{Complete form of polynomial \texorpdfstring{$P$ and $Q$}{Lg}}
\label{completepoly}

\begin{widetext}

Solving for $\pi$ from Eq.~(\ref{rhoa-pi-stat}) we obtain
\begin{equation}
  \pi = \frac{\beta \eta \rho_i^* \rho_{su}^* -\mu\rho_{iu}^*}{\rho_{su}^*(\beta \eta \rho_i^*  - \gamma \eta \rho_a^* ) -\mu\rho_{iu}^*+\alpha \rho_{sa}^*},
  \label{pi-rhoa}    
\end{equation}
which, after inserting expressions for $\rho_{iu}^*$, $\rho_i^*$, $\rho_{su}^*$ and $\rho_{sa}^*$ from Eqs.~(\ref{riu}), (\ref{ri-1}), (\ref{rsu-1}) and (\ref{rsa-1}), respectively, becomes
\begin{equation}
  \pi = \frac{P(\rho_a^*)}{Q(\rho_a^*)},
\end{equation}
with 
\begin{equation}
P(\rho_a^*) = \frac{\beta \eta}{\Gamma} \left[\frac{\beta \eta - \mu}{\beta \eta}-(1-\Gamma)\left(\rho_a^* +\frac{\left[ \alpha - \gamma \eta (1-\rho_a^*) \right] \rho_a^*}{\kappa} \right)\right]-\frac{\mu \left[ \alpha - \gamma \eta (1-\rho_a^*) \right] \rho_a^*}{\kappa},   
\label{P}
\end{equation}
and 
\begin{eqnarray}
 Q(\rho_a^*) &=& \left(1-\rho_a^*+\frac{\left[ \alpha - \gamma \eta (1-\rho_a^*) \right] \rho_a^*}{\kappa} \right )\left\{\frac{\beta \eta}{\Gamma} \left[\frac{\beta\eta-\mu}{\beta\eta}-(1-\Gamma)\left(\rho_a^*+\frac{\left[ \alpha - \gamma \eta (1-\rho_a^*) \right] \rho_a^*}{\kappa} \right)\right] -\gamma \eta \rho_a^*\right\}  \nonumber \\ 
 &-& \frac{\mu \left[ \alpha - \gamma \eta (1-\rho_a^*) \right] \rho_a^*}{\kappa} +\frac{\alpha}{\Gamma}\left[\rho_a^*+\frac{\Gamma\left[ \alpha - \gamma \eta (1-\rho_a^*) \right] \rho_a^*}{\kappa^2}-\frac{\beta\eta-\mu}{\beta\eta}\right].
\label{Q}
\end{eqnarray}

\end{widetext}

\section{Solution for \texorpdfstring{$\gamma=0 ~ \mbox{and} ~ \Gamma=0$}{Lg}}
\label{gG0}

For $\gamma=0$ and $\Gamma=0$ Eqs.~(\ref{drhodt}) are reduced to the simpler form  
\begin{subequations} 	
  \begin{alignat}{4}
   \label{drhoiudt-2}
\frac{d\rho_{iu}}{dt} &= (1-\pi) \beta \eta \rho_{su} \rho_{i} + \pi
\alpha \rho_{ia} - (1-\pi) \mu \rho_{iu} - \pi \rho_{iu},  \\
\label{drhosudt-2}
\frac{d\rho_{su}}{dt} &= (1-\pi) \mu \rho_{iu} + \pi \alpha \rho_{sa}
- (1-\pi) \beta \eta \rho_{su} \rho_i, \\
\label{drhoiadt-2}
\frac{d\rho_{ia}}{dt} &= \pi \rho_{iu} - \pi \alpha \rho_{ia} - (1-\pi) \mu
\rho_{ia}, \\
\label{drhosadt-2}
\frac{d\rho_{sa}}{dt} &= (1-\pi) \mu \rho_{ia} - \pi \alpha \rho_{sa}. 
  \end{alignat}
  \label{drhodt-2}
\end{subequations} 
 
The trivial fixed point of this system of equations is
$\rho_{su}^*=1.0$, corresponding to a totally healthy and unaware
population. The non-trivial fixed point corresponds to the stationary
densities
\begin{eqnarray}  
\rho_{iu}^* &=& \frac{\alpha (\beta \eta - \mu)}{(\kappa+\alpha)\beta \eta} \\
\rho_{ia}^* &=& \frac{\pi \alpha \kappa (\beta \eta -\mu)}{(\kappa+\alpha) \beta \eta 
  \left[\pi \alpha +(1-\pi) \mu \right]} \\
\rho_{su}^* &=& \frac{\mu}{\beta \eta} \\
\rho_{sa}^* &=& \frac{(1-\pi) \mu \kappa (\beta \eta -\mu)}{(\kappa+\alpha) \beta \eta \left[ \pi \alpha + (1-\pi) \mu \right]}.
\end{eqnarray}
The expression for the disease prevalence is
\begin{equation}
\rho_{i}^* = \frac{\alpha (\beta \eta - \mu) \left[\pi (\kappa+\alpha) +
    (1-\pi)\mu \right]}{(\kappa+\alpha) \beta \eta \left[ \pi \alpha +
    (1-\pi) \mu \right]} 
\label{rhoi}
\end{equation}

Equation~(\ref{rhoi}) predicts that the prevalence takes the value 
$\rho_i^*=\alpha (\beta \eta - \mu)/[(1+\alpha) \beta \eta]=0.1875$ and 
$\rho_i^*=(\beta \eta - \mu)/\beta \eta=0.5$ in the $\pi=0$ and $\pi=1$
limits, respectively.  However, 
these extreme cases are pathological because the above limiting
values do not correspond to the value of $\rho_i^*$ at those points.
That is, $\rho_i^*$ exhibits a discontinuity at $\pi=0$ and at
$\pi=1$.  To see that we rewrite Eqs.~(\ref{drhodt-2}) for $\pi=0$
\begin{eqnarray}
\frac{d\rho_{iu}}{dt} &=& \beta \eta \rho_{su} \rho_{i} - \mu
\rho_{iu}, \nonumber \\
\label{drhodt2}
\frac{d\rho_{su}}{dt} &=& \mu \rho_{iu} - \beta \eta \rho_{su} \rho_i, \\
\frac{d\rho_{ia}}{dt} &=& - \mu \rho_{ia}, \nonumber \\
\frac{d\rho_{sa}}{dt} &=& \mu \rho_{ia}, \nonumber 
\end{eqnarray}
whose non-trivial stationary solution is $\rho_{iu}^*=C_0-\mu/\beta \eta$,
$\rho_{ia}^*=0$, $\rho_{su}^*=\mu/\beta \eta$ and $\rho_{sa}^*=1-C_0$, where
$C_0=\rho_u(t=0)$ is a constant.  Assuming that all individuals are
unaware initially, $C_0=1$, leads to a prevalence 
$\rho_i^*=(\beta \eta-\mu)/\beta \eta=0.5$ at $\pi=0$, which is higher
by a factor $(1+\alpha)/\alpha=2.66$ than the limit $\pi \to 0$ from
Eq.~(\ref{rhoi}). For $\pi=1$ Eqs.~(\ref{drhodt-2}) are reduced to
\begin{eqnarray}
\frac{d\rho_{iu}}{dt} &=& \alpha \rho_{ia} - \rho_{iu},  \nonumber \\
\label{drhodt3}
\frac{d\rho_{su}}{dt} &=& \alpha \rho_{sa}, \\
\frac{d\rho_{ia}}{dt} &=& \rho_{iu} - \alpha \rho_{ia}, \nonumber \\
\frac{d\rho_{sa}}{dt} &=& - \alpha \rho_{sa}, \nonumber 
\end{eqnarray}
whose stationary solution is $\rho_{iu}^*=\alpha C_1/(1+\alpha)$,
$\rho_{ia}^*=C_1/(1+\alpha)$, $\rho_{su}^*=1-C_1$ and $\rho_{sa}^*=0$, where
$C_1=\rho_i(t=0)$.  That is, the fraction of infected nodes stays
constant over time.  If there is one infected individual initially, then the
prevalence is $\rho_i^*=1/N \ll 1$ for large $N$.  

We note that the stationary density of aware nodes $\rho_a^*=(\beta \eta-\mu)/[\beta \eta(1+\alpha)]$ is independent on $\pi$, while $\rho_i^*$ does depend to $\pi$.  This means that both SIS and UAU dynamics are cyclic but not equivalent. This equivalence is broken by the term $\kappa \pi$ in the spontaneous transition $IU \to IA$.  Indeed, for the $\kappa=0$ case we obtain that $\rho_i^*=(\beta \eta - \mu)/\beta \eta$ independent on $\pi$.  This gives an insight into the non-intuitive behavior of $\rho_i^*$, as we describe in section~\ref{stationary}.

\section{Linear stability analysis}
\label{stability}

To better handle calculations, we write the fixed points of Eqs.~(\ref{rho1rho2}) in the general form  
$\vec{\rho}_n^{\, *}=(0,A,0,1-A)$, where 
\begin{eqnarray}
A &=& 0 ~~~~~~~~~~~ \mbox{for $\gamma \eta < \alpha$ ($n=1$) and} \nonumber \\ 
A &=& \frac{\gamma \eta -\alpha}{\gamma \eta} ~~~ \mbox{for $\gamma \eta > \alpha$ ($n=2$)}, 
\label{A1}
\end{eqnarray}
and study their stability under a small perturbation by means of Eqs.~(\ref{drhodt}).  For that, we linearize Eqs.~(\ref{drhodt}) around the fixed point $\vec{\rho}_n^{\, *}$ by setting
$\rho_{iu}=\epsilon_1$, $\rho_{sa}=A+\epsilon_2$ and $\rho_{ia}=\epsilon_3$, with $| \epsilon_k | \ll 1$ ($k=1,2,3$), and study their time evolution (the evolution of $\rho_{su}$ is obtained from the other three densities).  Neglecting terms of order $\epsilon_k^2$, we obtain
\begin{equation}
    \frac{d \vec{\epsilon}}{dt}= \mathbf{M} \, \vec{\epsilon}
\end{equation}
where 
\begin{eqnarray*}
\textbf{M} \equiv
\begin{pmatrix}
a & 0 & b \\
c & d & e \\
f & 0 & g 
\end{pmatrix} ~~~ \mbox{and} ~~~~
\vec{\epsilon} \equiv 
\begin{pmatrix}
\epsilon_1, \epsilon_2, \epsilon_3 \\
\end{pmatrix}, 
\end{eqnarray*}
with
\begin{eqnarray*}
a &=&  (1-\pi) \left[\beta \eta (1-A)-\mu\right]-\pi\left[\kappa+\eta\gamma A\right], \\
b &=&  (1-\pi) \beta \eta (1-A)+\pi\alpha, \\
c &=&  - \left[ \pi \gamma \eta + (1-\pi)\Gamma \beta \eta \right] A , \\
d &=&   \pi \left[ \gamma \eta (1- 2A) - \alpha \right] , \\
e &=&   \pi\gamma\eta (1- 2A) + (1-\pi) \left[ \mu - \Gamma\beta\eta A \right], \\
f &=&    \pi \left[ \gamma\eta A + \kappa \right] + (1-\pi)\Gamma \beta \eta A, \\
g &=&   (1-\pi) \left[ \Gamma\beta\eta A - \mu \right] - \pi\alpha.
\end{eqnarray*}
At the critical point, the determinant of matrix $\textbf{M}$ 
$$det(\textbf{M})=d(ag-fb)$$ 
must be zero, from where obtain after doing some algebra the relation quoted in Eq.~(\ref{relation}) of the main text.

\bibliographystyle{apsrev}

\bibliography{references}

\begin{thebibliography}{27}
\expandafter\ifx\csname natexlab\endcsname\relax\def\natexlab#1{#1}\fi
\expandafter\ifx\csname bibnamefont\endcsname\relax
  \def\bibnamefont#1{#1}\fi
\expandafter\ifx\csname bibfnamefont\endcsname\relax
  \def\bibfnamefont#1{#1}\fi
\expandafter\ifx\csname citenamefont\endcsname\relax
  \def\citenamefont#1{#1}\fi
\expandafter\ifx\csname url\endcsname\relax
  \def\url#1{\texttt{#1}}\fi
\expandafter\ifx\csname urlprefix\endcsname\relax\def\urlprefix{URL }\fi
\providecommand{\bibinfo}[2]{#2}
\providecommand{\eprint}[2][]{\url{#2}}

\bibitem[{\citenamefont{Pastor-Satorras
  et~al.}(2015)\citenamefont{Pastor-Satorras, Castellano, Van~Mieghem, and
  Vespignani}}]{Pastor-Satorras-2015}
\bibinfo{author}{\bibfnamefont{R.}~\bibnamefont{Pastor-Satorras}},
  \bibinfo{author}{\bibfnamefont{C.}~\bibnamefont{Castellano}},
  \bibinfo{author}{\bibfnamefont{P.}~\bibnamefont{Van~Mieghem}},
  \bibnamefont{and}
  \bibinfo{author}{\bibfnamefont{A.}~\bibnamefont{Vespignani}},
  \bibinfo{journal}{Rev. Mod. Phys.} \textbf{\bibinfo{volume}{87}},
  \bibinfo{pages}{925} (\bibinfo{year}{2015}).

\bibitem[{\citenamefont{Ferraz~de Arruda et~al.}(2018)\citenamefont{Ferraz~de
  Arruda, Rodrigues, and Moreno}}]{Arruda-2018}
\bibinfo{author}{\bibfnamefont{G.}~\bibnamefont{Ferraz~de Arruda}},
  \bibinfo{author}{\bibfnamefont{F.~A.} \bibnamefont{Rodrigues}},
  \bibnamefont{and} \bibinfo{author}{\bibfnamefont{Y.}~\bibnamefont{Moreno}},
  \bibinfo{journal}{Physics Reports} \textbf{\bibinfo{volume}{756}},
  \bibinfo{pages}{1} (\bibinfo{year}{2018}).

\bibitem[{\citenamefont{Wang et~al.}(2016{\natexlab{a}})\citenamefont{Wang,
  Bauch, Bhattacharyya, d'Onofrio, Manfredi, Perc, Perra, Salath{\'e}, and
  Zhao}}]{Wang016}
\bibinfo{author}{\bibfnamefont{Z.}~\bibnamefont{Wang}},
  \bibinfo{author}{\bibfnamefont{C.~T.} \bibnamefont{Bauch}},
  \bibinfo{author}{\bibfnamefont{S.}~\bibnamefont{Bhattacharyya}},
  \bibinfo{author}{\bibfnamefont{A.}~\bibnamefont{d'Onofrio}},
  \bibinfo{author}{\bibfnamefont{P.}~\bibnamefont{Manfredi}},
  \bibinfo{author}{\bibfnamefont{M.}~\bibnamefont{Perc}},
  \bibinfo{author}{\bibfnamefont{N.}~\bibnamefont{Perra}},
  \bibinfo{author}{\bibfnamefont{M.}~\bibnamefont{Salath{\'e}}},
  \bibnamefont{and} \bibinfo{author}{\bibfnamefont{D.}~\bibnamefont{Zhao}},
  \bibinfo{journal}{Physics Reports} \textbf{\bibinfo{volume}{664}},
  \bibinfo{pages}{1} (\bibinfo{year}{2016}{\natexlab{a}}).

\bibitem[{\citenamefont{Lima et~al.}(2015)\citenamefont{Lima, De~Domenico,
  Pejovic, and Musolesi}}]{Lima-2015}
\bibinfo{author}{\bibfnamefont{A.}~\bibnamefont{Lima}},
  \bibinfo{author}{\bibfnamefont{M.}~\bibnamefont{De~Domenico}},
  \bibinfo{author}{\bibfnamefont{V.}~\bibnamefont{Pejovic}}, \bibnamefont{and}
  \bibinfo{author}{\bibfnamefont{M.}~\bibnamefont{Musolesi}},
  \bibinfo{journal}{Scientific Reports} \textbf{\bibinfo{volume}{5}},
  \bibinfo{pages}{10650} (\bibinfo{year}{2015}).

\bibitem[{\citenamefont{Funk et~al.}(2010)\citenamefont{Funk, Salath{\'e}, and
  Jansen}}]{funk2010modelling}
\bibinfo{author}{\bibfnamefont{S.}~\bibnamefont{Funk}},
  \bibinfo{author}{\bibfnamefont{M.}~\bibnamefont{Salath{\'e}}},
  \bibnamefont{and} \bibinfo{author}{\bibfnamefont{V.~A.}
  \bibnamefont{Jansen}}, \bibinfo{journal}{Journal of the Royal Society
  Interface} p. \bibinfo{pages}{rsif20100142} (\bibinfo{year}{2010}).

\bibitem[{\citenamefont{Manfredi and D'Onofrio}(2013)}]{manfredi2013modeling}
\bibinfo{author}{\bibfnamefont{P.}~\bibnamefont{Manfredi}} \bibnamefont{and}
  \bibinfo{author}{\bibfnamefont{A.}~\bibnamefont{D'Onofrio}},
  \emph{\bibinfo{title}{Modeling the interplay between human behavior and the
  spread of infectious diseases}} (\bibinfo{publisher}{Springer Science \&
  Business Media}, \bibinfo{year}{2013}).

\bibitem[{\citenamefont{Meloni et~al.}(2011)\citenamefont{Meloni, Perra,
  Arenas, G\'omez, Moreno, and Vespignani}}]{Meloni011}
\bibinfo{author}{\bibfnamefont{S.}~\bibnamefont{Meloni}},
  \bibinfo{author}{\bibfnamefont{N.}~\bibnamefont{Perra}},
  \bibinfo{author}{\bibfnamefont{A.}~\bibnamefont{Arenas}},
  \bibinfo{author}{\bibfnamefont{S.}~\bibnamefont{G\'omez}},
  \bibinfo{author}{\bibfnamefont{Y.}~\bibnamefont{Moreno}}, \bibnamefont{and}
  \bibinfo{author}{\bibfnamefont{A.}~\bibnamefont{Vespignani}},
  \bibinfo{journal}{Scientific Reports} \textbf{\bibinfo{volume}{1}}
  (\bibinfo{year}{2011}).

\bibitem[{\citenamefont{Wang et~al.}(2015)\citenamefont{Wang, Andrews, Wu,
  Wang, and Bauch}}]{Wang-2015-1}
\bibinfo{author}{\bibfnamefont{Z.}~\bibnamefont{Wang}},
  \bibinfo{author}{\bibfnamefont{M.~A.} \bibnamefont{Andrews}},
  \bibinfo{author}{\bibfnamefont{Z.-X.} \bibnamefont{Wu}},
  \bibinfo{author}{\bibfnamefont{L.}~\bibnamefont{Wang}}, \bibnamefont{and}
  \bibinfo{author}{\bibfnamefont{C.~T.} \bibnamefont{Bauch}},
  \bibinfo{journal}{Physics of Life Reviews} \textbf{\bibinfo{volume}{15}},
  \bibinfo{pages}{1 } (\bibinfo{year}{2015}).

\bibitem[{\citenamefont{Funka et~al.}(2015)\citenamefont{Funka, Bansal,
  T.Bauch, T.D.Eames, Edmunds, Galvani, and Klepac}}]{Funk_15}
\bibinfo{author}{\bibfnamefont{S.}~\bibnamefont{Funka}},
  \bibinfo{author}{\bibfnamefont{S.}~\bibnamefont{Bansal}},
  \bibinfo{author}{\bibfnamefont{C.}~\bibnamefont{T.Bauch}},
  \bibinfo{author}{\bibfnamefont{K.}~\bibnamefont{T.D.Eames}},
  \bibinfo{author}{\bibfnamefont{W.~J.} \bibnamefont{Edmunds}},
  \bibinfo{author}{\bibfnamefont{A.~P.} \bibnamefont{Galvani}},
  \bibnamefont{and} \bibinfo{author}{\bibfnamefont{P.}~\bibnamefont{Klepac}},
  \bibinfo{journal}{Epidemics} \textbf{\bibinfo{volume}{10}},
  \bibinfo{pages}{21} (\bibinfo{year}{2015}).

\bibitem[{\citenamefont{Funk et~al.}(2009)\citenamefont{Funk, Gilad, Watkins,
  and Jansen}}]{Funk6872}
\bibinfo{author}{\bibfnamefont{S.}~\bibnamefont{Funk}},
  \bibinfo{author}{\bibfnamefont{E.}~\bibnamefont{Gilad}},
  \bibinfo{author}{\bibfnamefont{C.}~\bibnamefont{Watkins}}, \bibnamefont{and}
  \bibinfo{author}{\bibfnamefont{V.~A.~A.} \bibnamefont{Jansen}},
  \bibinfo{journal}{Proceedings of the National Academy of Sciences}
  \textbf{\bibinfo{volume}{106}}, \bibinfo{pages}{6872} (\bibinfo{year}{2009}).

\bibitem[{\citenamefont{Granell et~al.}(2013)\citenamefont{Granell, G\'omez,
  and Arenas}}]{Granell-2013}
\bibinfo{author}{\bibfnamefont{C.}~\bibnamefont{Granell}},
  \bibinfo{author}{\bibfnamefont{S.}~\bibnamefont{G\'omez}}, \bibnamefont{and}
  \bibinfo{author}{\bibfnamefont{A.}~\bibnamefont{Arenas}},
  \bibinfo{journal}{Phys. Rev. Lett.} \textbf{\bibinfo{volume}{111}},
  \bibinfo{pages}{128701} (\bibinfo{year}{2013}).

\bibitem[{\citenamefont{Granell et~al.}(2014)\citenamefont{Granell, G{\'o}mez,
  and Arenas}}]{Granell-2014}
\bibinfo{author}{\bibfnamefont{C.}~\bibnamefont{Granell}},
  \bibinfo{author}{\bibfnamefont{S.}~\bibnamefont{G{\'o}mez}},
  \bibnamefont{and} \bibinfo{author}{\bibfnamefont{A.}~\bibnamefont{Arenas}},
  \bibinfo{journal}{Physical Review E} \textbf{\bibinfo{volume}{90}},
  \bibinfo{pages}{012808} (\bibinfo{year}{2014}).

\bibitem[{\citenamefont{Goldenberg et~al.}(2005)\citenamefont{Goldenberg,
  Shavitt, Shir, and Solomon}}]{Goldenberg-2005}
\bibinfo{author}{\bibfnamefont{J.}~\bibnamefont{Goldenberg}},
  \bibinfo{author}{\bibfnamefont{Y.}~\bibnamefont{Shavitt}},
  \bibinfo{author}{\bibfnamefont{E.}~\bibnamefont{Shir}}, \bibnamefont{and}
  \bibinfo{author}{\bibfnamefont{S.}~\bibnamefont{Solomon}},
  \bibinfo{journal}{Nature Physics} \textbf{\bibinfo{volume}{1}},
  \bibinfo{pages}{184} (\bibinfo{year}{2005}).

\bibitem[{\citenamefont{Newman}(2005)}]{Newman-2005}
\bibinfo{author}{\bibfnamefont{M.~E.~J.} \bibnamefont{Newman}},
  \bibinfo{journal}{Phys. Rev. Lett.} \textbf{\bibinfo{volume}{95}},
  \bibinfo{pages}{108701} (\bibinfo{year}{2005}).

\bibitem[{\citenamefont{Karrer and Newman}(2011)}]{Karrer-2011}
\bibinfo{author}{\bibfnamefont{B.}~\bibnamefont{Karrer}} \bibnamefont{and}
  \bibinfo{author}{\bibfnamefont{M.~E.~J.} \bibnamefont{Newman}},
  \bibinfo{journal}{Phys. Rev. E} \textbf{\bibinfo{volume}{84}},
  \bibinfo{pages}{036106} (\bibinfo{year}{2011}).

\bibitem[{\citenamefont{Marceau et~al.}(2011)\citenamefont{Marceau, No\"el,
  H\'ebert-Dufresne, Allard, and Dub\'e}}]{Marceau-2011}
\bibinfo{author}{\bibfnamefont{V.}~\bibnamefont{Marceau}},
  \bibinfo{author}{\bibfnamefont{P.-A.} \bibnamefont{No\"el}},
  \bibinfo{author}{\bibfnamefont{L.}~\bibnamefont{H\'ebert-Dufresne}},
  \bibinfo{author}{\bibfnamefont{A.}~\bibnamefont{Allard}}, \bibnamefont{and}
  \bibinfo{author}{\bibfnamefont{L.~J.} \bibnamefont{Dub\'e}},
  \bibinfo{journal}{Phys. Rev. E} \textbf{\bibinfo{volume}{84}},
  \bibinfo{pages}{026105} (\bibinfo{year}{2011}).

\bibitem[{\citenamefont{Jo et~al.}(2006)\citenamefont{Jo, Baek, and
  Moon}}]{Hang-Hyun-2006}
\bibinfo{author}{\bibfnamefont{H.-H.} \bibnamefont{Jo}},
  \bibinfo{author}{\bibfnamefont{S.~K.} \bibnamefont{Baek}}, \bibnamefont{and}
  \bibinfo{author}{\bibfnamefont{H.-T.} \bibnamefont{Moon}},
  \bibinfo{journal}{Physica A: Statistical Mechanics and its Applications}
  \textbf{\bibinfo{volume}{361}}, \bibinfo{pages}{534 } (\bibinfo{year}{2006}),
  ISSN \bibinfo{issn}{0378-4371}.

\bibitem[{\citenamefont{Wang et~al.}(2014)\citenamefont{Wang, Tang, Yang, Do,
  Lai, and Lee}}]{Wang-2014}
\bibinfo{author}{\bibfnamefont{W.}~\bibnamefont{Wang}},
  \bibinfo{author}{\bibfnamefont{M.}~\bibnamefont{Tang}},
  \bibinfo{author}{\bibfnamefont{H.}~\bibnamefont{Yang}},
  \bibinfo{author}{\bibfnamefont{Y.}~\bibnamefont{Do}},
  \bibinfo{author}{\bibfnamefont{Y.-C.} \bibnamefont{Lai}}, \bibnamefont{and}
  \bibinfo{author}{\bibfnamefont{G.}~\bibnamefont{Lee}},
  \bibinfo{journal}{Scientific Reports} \textbf{\bibinfo{volume}{4}},
  \bibinfo{pages}{5097} (\bibinfo{year}{2014}).

\bibitem[{\citenamefont{Wang et~al.}(2016{\natexlab{b}})\citenamefont{Wang,
  Liu, Cai, Braunstein, and Stanley}}]{Wang-2016}
\bibinfo{author}{\bibfnamefont{W.}~\bibnamefont{Wang}},
  \bibinfo{author}{\bibfnamefont{Q.~H.} \bibnamefont{Liu}},
  \bibinfo{author}{\bibfnamefont{M.}~\bibnamefont{Cai}, \bibfnamefont{S.~M.
  and.~Tang}}, \bibinfo{author}{\bibfnamefont{L.~A.} \bibnamefont{Braunstein}},
  \bibnamefont{and} \bibinfo{author}{\bibfnamefont{H.~E.}
  \bibnamefont{Stanley}}, \bibinfo{journal}{Scientific Reports}
  \textbf{\bibinfo{volume}{6}}, \bibinfo{pages}{29259}
  (\bibinfo{year}{2016}{\natexlab{b}}).

\bibitem[{\citenamefont{Guo et~al.}(2015)\citenamefont{Guo, Jiang, Lei, Li, Ma,
  and Zheng}}]{guo2015cascade}
\bibinfo{author}{\bibfnamefont{Q.}~\bibnamefont{Guo}},
  \bibinfo{author}{\bibfnamefont{X.}~\bibnamefont{Jiang}},
  \bibinfo{author}{\bibfnamefont{Y.}~\bibnamefont{Lei}},
  \bibinfo{author}{\bibfnamefont{M.}~\bibnamefont{Li}},
  \bibinfo{author}{\bibfnamefont{Y.}~\bibnamefont{Ma}}, \bibnamefont{and}
  \bibinfo{author}{\bibfnamefont{Z.}~\bibnamefont{Zheng}},
  \bibinfo{journal}{Physical Review E} \textbf{\bibinfo{volume}{91}},
  \bibinfo{pages}{012822} (\bibinfo{year}{2015}).

\bibitem[{\citenamefont{Wu et~al.}(2012)\citenamefont{Wu, Fu, Small, and
  Xu}}]{wu_2012}
\bibinfo{author}{\bibfnamefont{Q.}~\bibnamefont{Wu}},
  \bibinfo{author}{\bibfnamefont{X.}~\bibnamefont{Fu}},
  \bibinfo{author}{\bibfnamefont{M.}~\bibnamefont{Small}}, \bibnamefont{and}
  \bibinfo{author}{\bibfnamefont{X.-J.} \bibnamefont{Xu}},
  \bibinfo{journal}{Chaos: An Interdisciplinary Journal of Nonlinear Science}
  \textbf{\bibinfo{volume}{22}}, \bibinfo{pages}{013101}
  (\bibinfo{year}{2012}).

\bibitem[{\citenamefont{Vel\'asquez-Rojas and Vazquez}(2017)}]{FVR-2017}
\bibinfo{author}{\bibfnamefont{F.}~\bibnamefont{Vel\'asquez-Rojas}}
  \bibnamefont{and} \bibinfo{author}{\bibfnamefont{F.}~\bibnamefont{Vazquez}},
  \bibinfo{journal}{Physical Review E} \textbf{\bibinfo{volume}{95}},
  \bibinfo{pages}{052315} (\bibinfo{year}{2017}).

\bibitem[{\citenamefont{da~Silva et~al.}(2019)\citenamefont{da~Silva,
  Vel\'asquez-Rojas, Connaughton, Vazquez, Moreno, and Rodrigues}}]{Col-Br-1}
\bibinfo{author}{\bibfnamefont{P.~C.~V.} \bibnamefont{da~Silva}},
  \bibinfo{author}{\bibfnamefont{F.}~\bibnamefont{Vel\'asquez-Rojas}},
  \bibinfo{author}{\bibfnamefont{C.}~\bibnamefont{Connaughton}},
  \bibinfo{author}{\bibfnamefont{F.}~\bibnamefont{Vazquez}},
  \bibinfo{author}{\bibfnamefont{Y.}~\bibnamefont{Moreno}}, \bibnamefont{and}
  \bibinfo{author}{\bibfnamefont{F.~A.} \bibnamefont{Rodrigues}},
  \bibinfo{journal}{Physical Review E} \textbf{\bibinfo{volume}{100}},
  \bibinfo{pages}{032313} (\bibinfo{year}{2019}).

\bibitem[{\citenamefont{Maki and Thompson}(1973)}]{Maki}
\bibinfo{author}{\bibfnamefont{D.~P.} \bibnamefont{Maki}} \bibnamefont{and}
  \bibinfo{author}{\bibfnamefont{j.~a.} \bibnamefont{Thompson},
  \bibfnamefont{Maynard}}, \emph{\bibinfo{title}{Mathematical models and
  applications : with emphasis on the social, life, and management sciences}}
  (\bibinfo{publisher}{Englewood Cliffs, N.J. : Prentice-Hall},
  \bibinfo{year}{1973}), ISBN \bibinfo{isbn}{0135616700},
  \bibinfo{note}{includes bibliographies}.

\bibitem[{\citenamefont{Bianconi}(2018)}]{Bianconi-2018}
\bibinfo{author}{\bibfnamefont{G.}~\bibnamefont{Bianconi}},
  \emph{\bibinfo{title}{Multilayer networks: structure and function}}
  (\bibinfo{publisher}{Oxford University Press}, \bibinfo{year}{2018}).

\bibitem[{\citenamefont{Hoang et~al.}(2019)\citenamefont{Hoang, Coletti,
  Melegaro, Wallinga, Grijalva, Edmunds, Beutels, and Hens}}]{Hoang-2019}
\bibinfo{author}{\bibfnamefont{T.}~\bibnamefont{Hoang}},
  \bibinfo{author}{\bibfnamefont{P.}~\bibnamefont{Coletti}},
  \bibinfo{author}{\bibfnamefont{A.}~\bibnamefont{Melegaro}},
  \bibinfo{author}{\bibfnamefont{J.}~\bibnamefont{Wallinga}},
  \bibinfo{author}{\bibfnamefont{C.~G.} \bibnamefont{Grijalva}},
  \bibinfo{author}{\bibfnamefont{J.~W.} \bibnamefont{Edmunds}},
  \bibinfo{author}{\bibfnamefont{P.}~\bibnamefont{Beutels}}, \bibnamefont{and}
  \bibinfo{author}{\bibfnamefont{N.}~\bibnamefont{Hens}},
  \bibinfo{journal}{Epidemiology} \textbf{\bibinfo{volume}{30}}
  (\bibinfo{year}{2019}).

\bibitem[{\citenamefont{Ferreira et~al.}(2012)\citenamefont{Ferreira,
  Castellano, and Pastor-Satorras}}]{Ferreira-2012-epidemic}
\bibinfo{author}{\bibfnamefont{S.~C.} \bibnamefont{Ferreira}},
  \bibinfo{author}{\bibfnamefont{C.}~\bibnamefont{Castellano}},
  \bibnamefont{and}
  \bibinfo{author}{\bibfnamefont{R.}~\bibnamefont{Pastor-Satorras}},
  \bibinfo{journal}{Physical Review E} \textbf{\bibinfo{volume}{86}},
  \bibinfo{pages}{041125} (\bibinfo{year}{2012}).

\end{thebibliography}

\end{document}